\documentclass{article}
\usepackage[utf8]{inputenc} 
\usepackage[left=3.8cm,right=3.8cm,top=3.8cm,bottom=3.8cm]{geometry}
\usepackage{authblk}
\usepackage{amssymb,amsmath} 
\usepackage{graphicx} 
\usepackage{array} 
\usepackage{eurosym} 
\usepackage{xcolor} 
\usepackage{url}
\usepackage[ruled,linesnumbered,vlined]{algorithm2e} 
\usepackage{hyperref}

\usepackage{tabularx}
\newcolumntype{L}[1]{>{\raggedright\arraybackslash}p{#1}} 
\newcolumntype{C}[1]{>{\centering\arraybackslash}p{#1}} 
\newcolumntype{R}[1]{>{\raggedleft\arraybackslash}p{#1}} 
\usepackage{colortbl}
\usepackage{multirow}
\usepackage{rotating}

\usepackage{pifont}
\newcommand{\leadingzero}[1]{\ifnum #1<10 0\the#1\else\the#1\fi} 

\definecolor{darkgrey}{RGB}{28,28,28} 
\definecolor{mediumgrey}{RGB}{71,71,71} 
\definecolor{lightgrey}{RGB}{115,115,115} 
\definecolor{lightlightgrey}{RGB}{206,206,206} 
\definecolor{darkblue}{RGB}{79,101,140} 
\definecolor{mediumblue}{RGB}{112,144,200} 
\definecolor{lightblue}{RGB}{140,180,250} 
\definecolor{darkgreen}{RGB}{41,115,46} 
\definecolor{mediumgreen}{RGB}{65,180,73} 
\definecolor{lightgreen}{RGB}{90,250,101} 
\definecolor{mediumyellow}{RGB}{247,203,56} 
\definecolor{mediumorange}{RGB}{255,138,60} 
\definecolor{mediumred}{RGB}{219,73,55} 
\definecolor{mediumviolet}{RGB}{146,90,199} 
\definecolor{mediumturquoise}{RGB}{87,189,227} 

\newcommand{\mIndexTime}{t}

\newcommand{\mIndexBucket}{b}
\newcommand{\mSetBuckets}{\mathcal{B}}
\newcommand{\mIndexStation}{m}
\newcommand{\mSetStations}{\mathcal{M}}

\newcommand{\mSetOStations}{\mSetStations^{O}}

\newcommand{\mIndexWaypoint}{w}
\newcommand{\mSetWaypoints}{\mathcal{W}}

\newcommand{\mSetBucketParkingSpots}{{\mSetWaypoints^{SL}}}

\newcommand{\mIndexItemDescription}{d}
\newcommand{\mSetItemDescriptions}{\mathcal{D}}

\newcommand{\mMetricParPodSpeedWeight}{w^S}
\newcommand{\mMetricParPodUtilityWeight}{w^U}
\newcommand{\mMetricStorageLocationProminenceNoParam}{F^{SL}}
\newcommand{\mMetricPodSpeedNoParam}{F^{PS}}
\newcommand{\mMetricPodUtilityNoParam}{F^{PU}}
\newcommand{\mMetricPodCombinedScoreNoParam}{F^{PC}}
\newcommand{\mMetricStorageLocationProminence}[1]{\mMetricStorageLocationProminenceNoParam \left( #1 \right)}
\newcommand{\mMetricPodSpeed}[2]{\mMetricPodSpeedNoParam \left( #1 , #2 \right)}
\newcommand{\mMetricPodUtility}[2]{\mMetricPodUtilityNoParam \left( #1 , #2 \right)}
\newcommand{\mMetricPodCombinedScore}[2]{\mMetricPodCombinedScoreNoParam \left( #1 , #2 \right)}

\newcommand{\mParTimeOStationPickItemConstant}{T^{P}}

\newcommand{\mGetPodItemCountContainedNoParam}{f^C}
\newcommand{\mGetPodItemCountContained}[3]{\mGetPodItemCountContainedNoParam \left( #1, #2, #3 \right)}
\newcommand{\mGetDemandOverallNoParam}{f^D}
\newcommand{\mGetDemandOverall}[2]{\mGetDemandOverallNoParam \left( #1 , #2 \right)}
\newcommand{\mGetItemFrequencyNoParam}{f^F}
\newcommand{\mGetItemFrequency}[1]{\mGetItemFrequencyNoParam \left( #1 \right)}
\newcommand{\mGetShortestPathTimeNoParam}{f^{A^*_t}}
\newcommand{\mGetShortestPathTime}[2]{\mGetShortestPathTimeNoParam \left( #1, #2 \right)}

\newcommand{\graphicsmaindir}{media}

\usepackage{ifthen}
\newboolean{preprint}
\setboolean{preprint}{true}

\usepackage{url}

\begin{document}
	
\title{Active repositioning of storage units in\\Robotic Mobile Fulfillment Systems}

\author[1]{Marius Merschformann\footnote{marius.merschformann@uni-paderborn.de}}
\affil[1]{Paderborn University, Paderborn, Germany}

\date{\today}
\maketitle


\begin{abstract}
In our work we focus on Robotic Mobile Fulfillment Systems in e-commerce distribution centers. These systems were designed to increase pick rates by employing mobile robots bringing movable storage units (so-called pods) to pick and replenishment stations as needed, and back to the storage area afterwards. One advantage of this approach is that repositioning of inventory can be done continuously, even during pick and replenishment operations. This is primarily accomplished by bringing a pod to a storage location different than the one it was fetched from, a process we call \textit{passive pod repositioning}. Additionally, this can be done by explicitly bringing a pod from one storage location to another, a process we call \textit{active pod repositioning}. In this work we introduce first mechanisms for the latter technique and conduct a simulation-based experiment to give first insights of their effect.

\end{abstract}

\section{Introduction}

In today's increasingly fast-paced e-commerce an efficient distribution center is one crucial element of the supply chain. Hence, new automated parts-to-picker systems have been introduced to increase throughput. One of them is the Robotic Mobile Fulfillment System (RMFS). In a RMFS mobile robots are used to bring rack-like storage units (so-called pods) to pick stations as required, thus, eliminating the need for the pickers to walk and search the inventory. A task which can take up to 70 \% of their time in traditional picker-to-parts systems (see~\cite{Tompkins.2010}). This concept was first introduced by~\cite{Wurman.2008} and an earlier simulation work by~\cite{Hazard.2006}. The first company implementing the concept at large scale was Kiva Systems, nowadays known as Amazon Robotics.\\
One of the features of RMFS is the continuous resorting of inventory, i.e. every time a pod is brought back to the storage area a different storage location may be used. While this potentially increases flexibility and adaptability, rarely used pods may block prominent storage locations, unless they are moved explicitly. This raises the question whether active repositioning of pods, i.e. picking up a pod and moving it to a different storage location, can be usefully applied to further increase the overall throughput of the system. In order to address this issue we focus on two approaches for active repositioning. First, we look at repositioning done in parallel while the system is constantly active and, second, we look at repositioning during system downtime (e.g. nightly down periods). While the assignment of storage locations to inventory is a well studied problem in warehousing (see \cite{Gu.2007}) the repositioning of inventory is typically not considered for other systems, because it is usually very expensive.


\section{Repositioning in RMFS}

\begin{figure}[tb]
	\centering
	\includegraphics[width=0.8\textwidth]{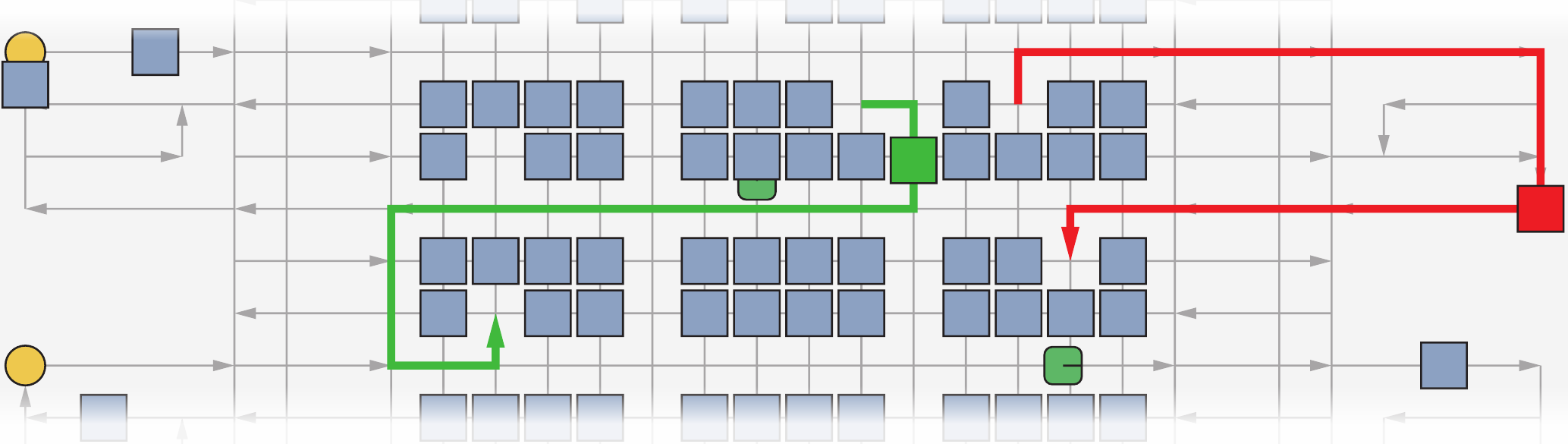}
	\caption{Active repositioning move (green arrow) vs. passive repositioning (red arrow)}
	\label{fig:activepassiverepositioning}
\end{figure}

In a RMFS \textit{passive repositioning} of pods is a natural process, if the storage location chosen for a pod is not fixed. For example, in many situations using the next available storage location is superior to a fixed strategy, because it decreases the travel time of the robots and by this enables an earlier availability for their next tasks. However, this strategy might cause no longer useful pods to be stored at very prominent storage locations, which introduces the blocking problem discussed earlier. There are two opportunities to resolve this issue: on the one hand it is possible to already consider characteristics of the pod content while choosing an appropriate storage location, while on the other hand it is possible to actively move pods from inappropriate storage locations to better fitting ones. We call the latter approach \textit{active repositioning} of pods. Both repositioning approaches are shown in Fig. \ref{fig:activepassiverepositioning}. Additionally, the figure shows an excerpt of the basic layout used for the experiments, i.e. the replenishment stations (yellow circles), pick stations (red circles), pods (blue squares), the storage locations (blocks of 2 by 4) and the directed waypoint graph used for path planning. This layout is based on the work by~\cite{Lamballais.2016}.

In order to assess the value of storage locations and pods we introduce the following metrics. First the \textit{prominence} $\mMetricStorageLocationProminenceNoParam$ of a storage location $\mIndexWaypoint \in \mSetBucketParkingSpots$ is determined by measuring the minimum shortest path time to a pick station $\mIndexStation \in \mSetOStations$ (see Eq. \ref{eq:storagelocationprominence}). The shortest path time $\mGetShortestPathTimeNoParam$ is computed with a modified A$^*$ algorithm that considers turning times to achieve more accurate results. The storage location with the lowest $\mMetricStorageLocationProminence{\mIndexWaypoint}$ is considered the most prominent one, since it offers the shortest time for bringing the pod to the next pick station. In order to assess the value of a pod $\mIndexBucket$ at time $\mIndexTime$ we introduce the pod-speed ($\mMetricPodSpeedNoParam$) and pod-utility ($\mMetricPodUtilityNoParam$) measures. The \textit{speed} of a pod (see Eq. \ref{eq:podspeed}) is calculated by summing up (across all SKUs) the units of an SKU contained ($\mGetPodItemCountContainedNoParam$) multiplied with the frequency of it ($\mGetItemFrequencyNoParam$). This frequency is a relative value reflecting the number of times a SKU is part of a customer order compared to all other SKUs. By using the minimum of units of an SKU contained and the demand for it ($\mGetDemandOverallNoParam$), the \textit{utility} of a pod (see Eq. \ref{eq:podutility}) sums the number of potential picks when considering the customer order backlog. Thus, it is a more dynamic value. Both scores are then combined in the metric $\mMetricPodCombinedScoreNoParam$ (see Eq. \ref{eq:podcombinedscore}). For our experiments we consider the weights $\mMetricParPodSpeedWeight = \mMetricParPodUtilityWeight = 1$ to value both characteristics equally.
\begin{align}
\label{eq:storagelocationprominence}
\mMetricStorageLocationProminence{\mIndexWaypoint} & := \min\limits_{\mIndexStation \in \mSetOStations} \mGetShortestPathTime{\mIndexWaypoint}{\mIndexStation} \\
\label{eq:podspeed}
\mMetricPodSpeed{\mIndexBucket}{\mIndexTime} & := \sum\limits_{\mIndexItemDescription \in \mSetItemDescriptions} \left( \mGetPodItemCountContained{\mIndexBucket}{\mIndexItemDescription
}{\mIndexTime} \cdot \mGetItemFrequency{\mIndexItemDescription} \right) \\
\label{eq:podutility}
\mMetricPodUtility{\mIndexBucket}{\mIndexTime} & := \sum\limits_{\mIndexItemDescription \in \mSetItemDescriptions} \left( \min \left( \mGetPodItemCountContained{\mIndexBucket}{\mIndexItemDescription
}{\mIndexTime} , \mGetDemandOverall{\mIndexItemDescription}{\mIndexTime} \right) \right) \\
\label{eq:podcombinedscore}
\mMetricPodCombinedScore{\mIndexBucket}{\mIndexTime} & := \frac{\mMetricPodSpeed{\mIndexBucket}{\mIndexTime}}{\max\limits_{\mIndexBucket ' \in \mSetBuckets} \mMetricPodSpeed{\mIndexBucket '}{\mIndexTime}} \cdot \mMetricParPodSpeedWeight + \frac{\mMetricPodUtility{\mIndexBucket}{\mIndexTime}}{\max\limits_{\mIndexBucket ' \in \mSetBuckets} \mMetricPodUtility{\mIndexBucket '}{\mIndexTime}} \cdot \mMetricParPodUtilityWeight
\end{align}
For evaluation purposes we can use these measures to determine an overall ``well-sortedness'' score for the inventory. The procedure for calculating the well-sortedness score is described in Alg. \ref{alg:wellsortedness}. At first we sort all storage locations by their prominence score in ascending order (see line \ref{line:storagelocationsorting} f.). Next, ranks $r_i$ are assigned to all storage locations $i \in \mSetBucketParkingSpots$, i.e., the best ones are assigned to the first rank and the rank is increased by one each time the prominence value increases (see line \ref{line:rankcalculation} f.). Then, we assess all storage location two-tuples and count misplacements, i.e., both storage locations are not of the same rank and the score of the better placed pod at $i_1$ is lower than the worse placed pod at $i_2$ (see line \ref{line:misplacementcalculation} f.). In addition to the number of misplacements we track the rank offset. From this we can calculate the average rank offset of all misplacements, i.e., the well-sortedness. Hence, a lower well-sortedness value means a better sorted inventory according to the given combined pod-speed and pod-utility measures.

\begin{algorithm}[tb]
	\caption{CalculateWellsortednessCombined}
	\label{alg:wellsortedness}
	\DontPrintSemicolon
	\LinesNumbered
	\SetVlineSkip{0mm}
	\SetInd{1mm}{2mm}
	\SetKwFunction{SortList}{Sort}
	\SetKwFunction{ArrayLen}{Size}
	\SetKwFunction{PodStored}{IsPodStored}
	\SetKwFunction{GetPod}{GetPod}
	
	$\mathcal{L} \leftarrow$ \SortList{ $\mSetBucketParkingSpots$, $i \Rightarrow \mMetricStorageLocationProminence{i}$ } , $r' \leftarrow 1$, $f' \leftarrow \min\limits_{i \in \mSetBucketParkingSpots} \mMetricStorageLocationProminence{i}$ \\ \label{line:storagelocationsorting}
	
	\ForEach{$i \in \{ 0, \dots , $ \ArrayLen{$\mathcal{L}$}$-1\}$}{\label{line:rankcalculation}
		\lIf{$\mMetricStorageLocationProminence{i} > f'$}{
			$r' \leftarrow r' + 1$, $f' \leftarrow \mMetricStorageLocationProminence{i}$
		}
		$r_i \leftarrow r'$
	}
	
	$c \leftarrow 0$, $d \leftarrow 0$\\ \label{line:misplacementcalculation}
	\ForEach{$i_1 \in \{ 0, \dots , $ \ArrayLen{$\mathcal{L}$}$-1\}$}{
		\ForEach{$i_2 \in \{ i_1 + 1, \dots , $ \ArrayLen{$\mathcal{L}$}$-1\}$}{
			\If{\PodStored{$\mathcal{L}[i_1]$} $\wedge$ \PodStored{$\mathcal{L}[i_2]$} $\wedge$ $r_{\mathcal{L}[i_1]} \neq r_{\mathcal{L}[i_2]}$}{
				$\mIndexBucket_1 \leftarrow$ \GetPod{$\mathcal{L}[i_1]$}, $\mIndexBucket_2 \leftarrow$ \GetPod{$\mathcal{L}[i_2]$}\\
				\lIf{$\mMetricPodCombinedScore{\mIndexBucket_1}{\mIndexTime} < \mMetricPodCombinedScore{\mIndexBucket_2}{\mIndexTime}$}{
					$c \leftarrow c + 1, d \leftarrow d + \left( r_{\mathcal{L}[i_2]} - r_{\mathcal{L}[i_1]} \right)$
				}
			}
		}
	}
	\Return $a \leftarrow \frac{d}{c}$\\
\end{algorithm}

In this work we investigate the following repositioning mechanisms:

\begin{description}
	\item[Nearest (N)] For passive repositioning this mechanism always uses the nearest available storage location in terms of estimated path time ($\mGetShortestPathTimeNoParam$). This mechanism does not allow active repositioning.
	\item[Cache (C)] This mechanism uses the nearest 25 \% of storage locations in terms of estimated path time ($\mGetShortestPathTimeNoParam$) as a cache. During passive repositioning pods with combined score ($\mMetricPodCombinedScoreNoParam$) above a determined threshold are stored at a cache storage location and others are stored at one of the remaining storage locations. In its active variant it swaps pods from and to the cache.
	\item[Utility (U)] This mechanism matches the pods with the ranks of the storage locations (see Alg. \ref{alg:wellsortedness}) on the basis of their combined score ($\mMetricPodCombinedScoreNoParam$). A nearby storage location with a close by rank is selected during passive repositioning. In its active variant pods with the largest difference between their desired and their actual storage location are moved to an improved one.
\end{description}


\section{Computational results}

For capturing and studying the behavior of RMFS we use an event-driven agent-based simulation that considers acceleration / deceleration and turning times of the robots (see~\cite{Merschformann.2017b}). Since diverse decision problems need to be considered in an RMFS we focus the scope of the work by fixing all remaining mandatory ones to simple assignment policies and the FAR path planning algorithm described in \cite{Merschformann.2017}. A more detailed overview of the core decision problems of our scope are given in~\cite{Merschformann.2017b}. For all experiments we consider a simulation horizon of one week, do 5 repetitions to reduce noise and new customer and replenishment orders are generated in a random stream with a Gamma distribution ($k = 1, \Theta = 2$) used for the choice of SKU per order line from 1000 possible SKUs. Furthermore, we analyze repositioning for four layouts. The specific characteristics are set as follows:
\vspace{-\baselineskip}
\begin{center}
\begin{tabular}{l|cccc}
	Layout & Small & Wide & Long & Large \\
	\hline
	Stations (pick / replenish) & 4/4 & 8/8 & 4/4 & 8/8 \\
	Aisles (hor. x vert.) & 8x10 & 16x10 & 8x22 & 16x22 \\
	Pods & 673 & 1271 & 1407 & 2658 \\
\end{tabular}
\end{center}
For the evaluation of active repositioning effectiveness we consider two scenarios. At first, we look at a situation where the system faces a nightly down period (22:00 - 6:00) during which no worker is available for picking or replenishment, but robots can be used for active repositioning. In order to keep the replenishment processes from obscuring the contribution of nightly inventory sorting, replenishment orders are submitted to the system at 16:00 in the afternoon in an amount that is sufficient to bring the storage utilization back to 75 \% fill level. For pick operations we keep a constant backlog of 2000 customer orders to keep the system under pressure. Additionally, we generate 1500 orders per station at 22:00 in the evening to increase information for the pod utility metric about the demands for the following day. Secondly, we look at active repositioning done in parallel in a system that is continuously in action. For this, we consider three subordinate configurations distributing the robots per station as following:
\vspace{-2mm}
\begin{description}
	\setlength\itemsep{0.5mm}
	\item[R1P3A0:] $\frac{1}{4}$ replenishment, $\frac{3}{4}$ picking, no active repos.
	\item[R1P2A1:] $\frac{1}{4}$ replenishment, $\frac{2}{4}$ picking, $\frac{1}{4}$ active repos.
	\item[R1P3A1:] $\frac{1}{5}$ replenishment, $\frac{3}{5}$ picking, $\frac{1}{5}$ active repos. (+1 robot per station)
\end{description}
This scenario is kept under continuous pressure by keeping a backlog of constant size for both: replenishment (200) and customer orders (2000).\\
The main performance metric for the evaluation is given by the unit throughput rate score (UTRS). Since we use a constant time of $\mParTimeOStationPickItemConstant = 10 s$ for picking one unit an upper bound for the number of units that can possibly be handled by the system during active hours can be calculated by $UB := \left| \mSetOStations \right| \frac{3600}{\mParTimeOStationPickItemConstant}$ with the set of all pick stations $\mSetOStations$. Using this we can determine the fractional score by dividing the actual picked units per hour in average by this upper bound.

\begin{table}[tb]
	\centering
	\setlength{\tabcolsep}{0.3mm} 
	\caption{Unit throughput rate score of the different scenarios, layouts and mechanisms (values in percent (\%), read columns as [passive mechanism-active mechanism])}
	\label{tab:computationalresults}
	\begin{tabular}{l|cccc|cccc|cccc|cccc}
		Layout & \multicolumn{4}{c}{Small} & \multicolumn{4}{c}{Wide} & \multicolumn{4}{c}{Long} & \multicolumn{4}{c}{Large}  \\
		Setup/Mech. & C-C & U-U & N-C & N-U & C-C & U-U & N-C & N-U & C-C & U-U & N-C & N-U & C-C & U-U & N-C & N-U \\
		\hline
		Deactivated & \cellcolor{mediumyellow!2.1!mediumred!80!white}52.9 & \cellcolor{mediumyellow!45.2!mediumred!80!white}53.6 & \cellcolor{mediumgreen!35.7!mediumyellow!80!white}55.0 & \cellcolor{mediumgreen!35.7!mediumyellow!80!white}55.0 & \cellcolor{mediumyellow!11.3!mediumred!80!white}50.6 & \cellcolor{mediumyellow!9.6!mediumred!80!white}50.5 & \cellcolor{mediumgreen!46.8!mediumyellow!80!white}52.7 & \cellcolor{mediumgreen!46.8!mediumyellow!80!white}52.7 & \cellcolor{mediumyellow!13.5!mediumred!80!white}43.3 & \cellcolor{mediumyellow!92.4!mediumred!80!white}46.0 & \cellcolor{mediumgreen!43.7!mediumyellow!80!white}47.8 & \cellcolor{mediumgreen!43.7!mediumyellow!80!white}47.8 & \cellcolor{mediumyellow!12.3!mediumred!80!white}42.6 & \cellcolor{mediumyellow!87.9!mediumred!80!white}44.8 & \cellcolor{mediumgreen!38.1!mediumyellow!80!white}46.3 & \cellcolor{mediumgreen!36.6!mediumyellow!80!white}46.3 \\
		Activated & \cellcolor{mediumyellow!0.2!mediumred!80!white}52.9 & \cellcolor{mediumyellow!39.4!mediumred!80!white}53.5 & \cellcolor{mediumgreen!69.6!mediumyellow!80!white}55.5 & \cellcolor{mediumgreen!99.8!mediumyellow!80!white}55.9 & \cellcolor{mediumyellow!8.9!mediumred!80!white}50.5 & \cellcolor{mediumyellow!0.2!mediumred!80!white}50.4 & \cellcolor{mediumgreen!74.1!mediumyellow!80!white}53.1 & \cellcolor{mediumgreen!99.8!mediumyellow!80!white}53.5 & \cellcolor{mediumyellow!0.2!mediumred!80!white}42.8 & \cellcolor{mediumgreen!1.2!mediumyellow!80!white}46.3 & \cellcolor{mediumgreen!81.2!mediumyellow!80!white}49.0 & \cellcolor{mediumgreen!99.8!mediumyellow!80!white}49.7 & \cellcolor{mediumyellow!0.2!mediumred!80!white}42.2 & \cellcolor{mediumyellow!97.3!mediumred!80!white}45.1 & \cellcolor{mediumgreen!81.1!mediumyellow!80!white}47.6 & \cellcolor{mediumgreen!99.8!mediumyellow!80!white}48.2 \\
		\hline
		R1P3A0 & \cellcolor{mediumgreen!59.4!mediumyellow!80!white}52.6 & \cellcolor{mediumgreen!47.4!mediumyellow!80!white}51.7 & \cellcolor{mediumgreen!99.8!mediumyellow!80!white}55.6 & \cellcolor{mediumgreen!99.8!mediumyellow!80!white}55.6 & \cellcolor{mediumgreen!56.3!mediumyellow!80!white}50.3 & \cellcolor{mediumgreen!38.2!mediumyellow!80!white}48.9 & \cellcolor{mediumgreen!95.4!mediumyellow!80!white}53.3 & \cellcolor{mediumgreen!95.3!mediumyellow!80!white}53.3 & \cellcolor{mediumgreen!31.4!mediumyellow!80!white}43.3 & \cellcolor{mediumgreen!32.5!mediumyellow!80!white}43.4 & \cellcolor{mediumgreen!87.5!mediumyellow!80!white}48.2 & \cellcolor{mediumgreen!87.7!mediumyellow!80!white}48.2 & \cellcolor{mediumgreen!25.4!mediumyellow!80!white}42.4 & \cellcolor{mediumgreen!25.6!mediumyellow!80!white}42.4 & \cellcolor{mediumgreen!74.4!mediumyellow!80!white}46.6 & \cellcolor{mediumgreen!74.5!mediumyellow!80!white}46.6 \\
		R1P2A1 & \cellcolor{mediumyellow!2.9!mediumred!80!white}40.9 & \cellcolor{mediumyellow!0.2!mediumred!80!white}40.7 & \cellcolor{mediumyellow!52.2!mediumred!80!white}44.6 & \cellcolor{mediumyellow!41.2!mediumred!80!white}43.8 & \cellcolor{mediumyellow!20.6!mediumred!80!white}39.7 & \cellcolor{mediumyellow!0.2!mediumred!80!white}38.1 & \cellcolor{mediumyellow!66.9!mediumred!80!white}43.3 & \cellcolor{mediumyellow!53.9!mediumred!80!white}42.3 & \cellcolor{mediumyellow!0.2!mediumred!80!white}31.7 & \cellcolor{mediumyellow!23.9!mediumred!80!white}33.8 & \cellcolor{mediumyellow!83.6!mediumred!80!white}39.1 & \cellcolor{mediumyellow!72.5!mediumred!80!white}38.1 & \cellcolor{mediumyellow!0.2!mediumred!80!white}31.5 & \cellcolor{mediumyellow!15.1!mediumred!80!white}32.8 & \cellcolor{mediumyellow!84.3!mediumred!80!white}38.8 & \cellcolor{mediumyellow!72.3!mediumred!80!white}37.8 \\
		R1P3A1 & \cellcolor{mediumgreen!50.7!mediumyellow!80!white}51.9 & \cellcolor{mediumgreen!44.8!mediumyellow!80!white}51.5 & \cellcolor{mediumgreen!96.4!mediumyellow!80!white}55.3 & \cellcolor{mediumgreen!85.2!mediumyellow!80!white}54.5 & \cellcolor{mediumgreen!56.6!mediumyellow!80!white}50.3 & \cellcolor{mediumgreen!33.4!mediumyellow!80!white}48.5 & \cellcolor{mediumgreen!99.8!mediumyellow!80!white}53.7 & \cellcolor{mediumgreen!91.6!mediumyellow!80!white}53.0 & \cellcolor{mediumgreen!13.0!mediumyellow!80!white}41.6 & \cellcolor{mediumgreen!39.2!mediumyellow!80!white}44.0 & \cellcolor{mediumgreen!99.8!mediumyellow!80!white}49.3 & \cellcolor{mediumgreen!86.0!mediumyellow!80!white}48.1 & \cellcolor{mediumgreen!11.2!mediumyellow!80!white}41.2 & \cellcolor{mediumgreen!27.8!mediumyellow!80!white}42.6 & \cellcolor{mediumgreen!99.8!mediumyellow!80!white}48.9 & \cellcolor{mediumgreen!85.9!mediumyellow!80!white}47.6 \\
	\end{tabular}
\end{table}

The results of the experiment are summarized in Tab. \ref{tab:computationalresults}. For the comparison of resorting the inventory during the nightly down period (line: Activated) vs. no active repositioning at all (line: Deactivated) we can observe an advantage in throughput. However, for the parallel active repositioning it is not possible to observe a positive effect. When moving one robot per pick station from pick operations to active repositioning (lines: R1P3A0 and R1P2A1) we observe a loss in UTRS, because less robots bring inventory to the pick stations. Even with an additional robot per pick station (line: R1P3A1) we cannot observe a substantial positive effect. For most cases, the effect is rather negative as a result of the increased congestion potential for robots moving within the storage area.

\begin{figure}[tb]
	\centering
	\includegraphics[width=\textwidth]{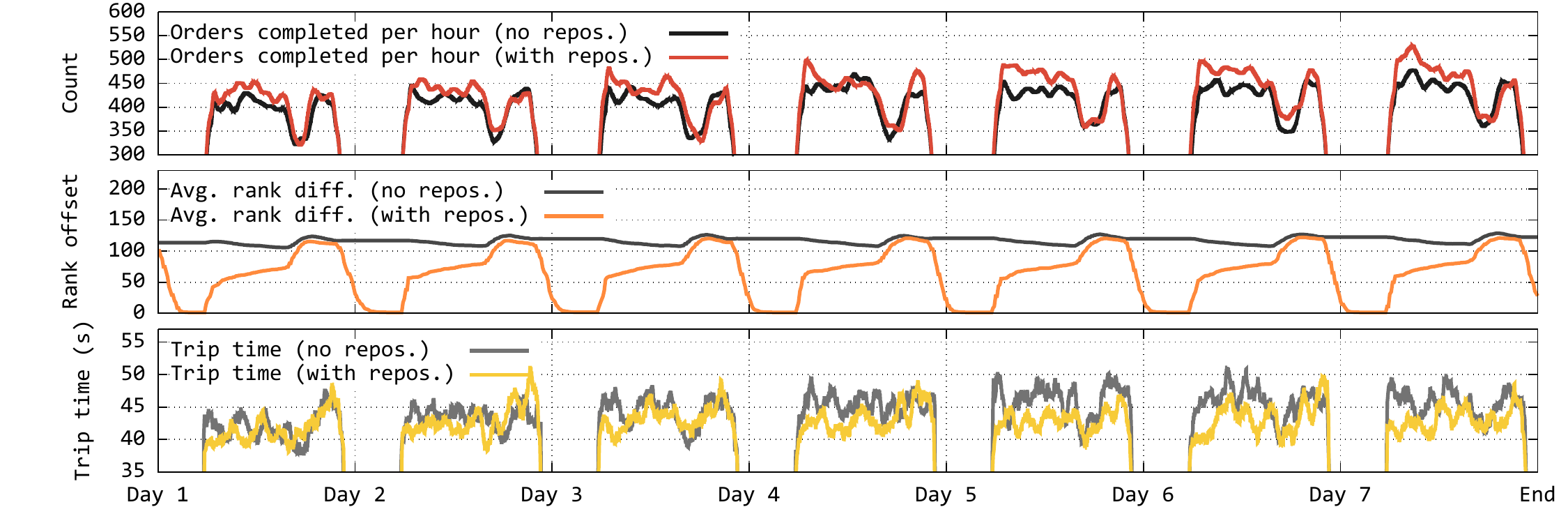}
	\caption{Time-wise comparison of layout Long and mechanisms N-U with (colored lines) and without (gray lines) active repositioning at night}
	\label{fig:comparisonovertime}
\end{figure}

In the following we take a closer look at the nightly down period scenario. If we keep the system sorted with the passive repositioning mechanism (C-C and U-U), nightly active repositioning does not have a noticeable positive effect, because the passive repositioning mechanism already keeps the inventory sorted for the most part. However, the Nearest mechanism which has a better overall performance, can benefit from a nightly active repositioning (N-C and N-U). Especially for the Large and Long layouts we can observe a reasonable boost in UTRS. The greater merit for layouts with more vertical aisles suggest that shorter trip times of the robots are the reason. This can also be observed when looking at the detailed results of a run with and without active repositioning (see Fig. \ref{fig:comparisonovertime}). First, more orders can be completed per hour after the inventory was sorted over night (first graph). This boost is eliminated as soon as replenishment operations begin. Thus, when and how replenishment is done is crucial to the benefit of resorting during down times, because the effect may be lost quite quickly. In the third graph the shorter times for completing trips to the pick stations after sorting the inventory support the assumption that these are the the main reason for the boost. Lastly, the second graph provides the well-sortedness measure and shows that sorting the inventory can be done reasonably fast.
\ifthenelse{\boolean{preprint}}{
	The situation before and after sorting is shown in Fig. \ref{fig:comparisonbyheatmap}. In this heatmap the combined score ($\mMetricPodCombinedScore{\mIndexBucket}{\mIndexTime}$) is visualized with one colored tile of the size of a storage location. Here the most useful and just replenished pods can be seen positioned next to the replenishment stations (left side) before sorting. After sorting most useful pods are positioned on the far right side of the horizontal aisles inbound to a pick station. Thus, these pods offering a high potential hit-rate (number of picks from a pod) can be fetched most quickly.\\
	
	\begin{figure}[tb]\textsl{}
		\centering
		\includegraphics[width=\textwidth]{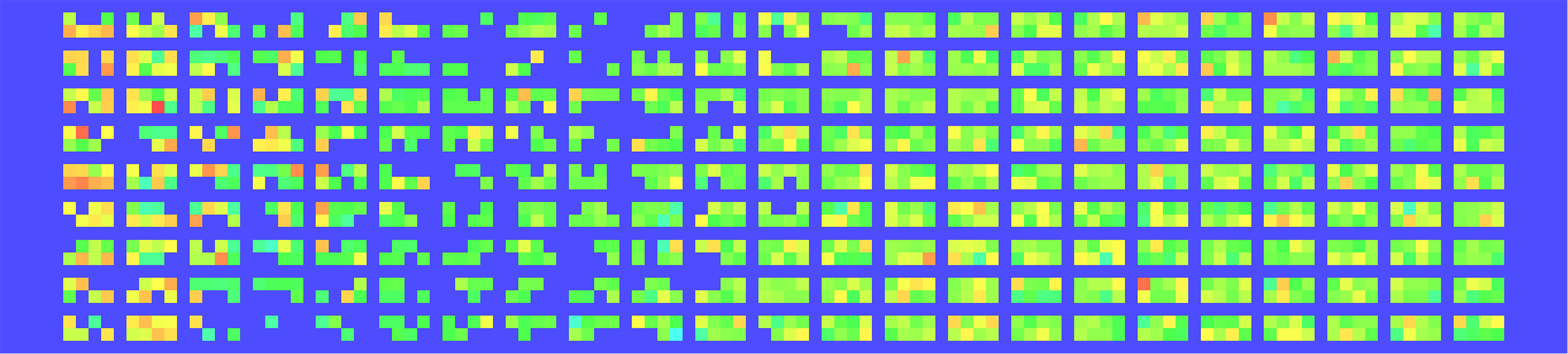}
		\includegraphics[width=\textwidth]{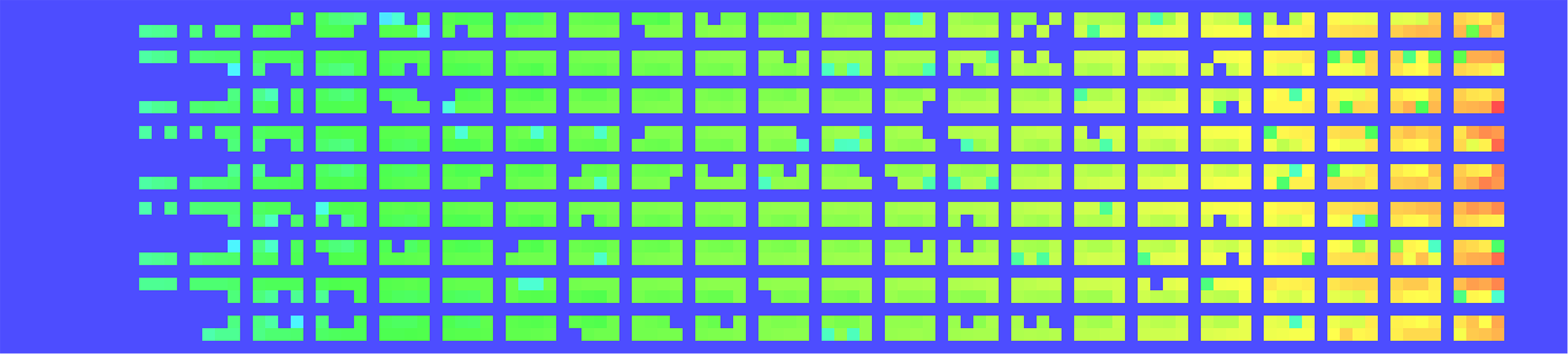}
		\caption{Comparison of inventory situation before and after nightly active repositioning (Long layout, N-U mechanisms, top: day 4 22:00, bottom: day 5 06:00)}
		\label{fig:comparisonbyheatmap}
	\end{figure}
}{}


\section{Conclusion}

The results suggest that active repositioning may boost throughput performance of RMFS. If the system faces regular down periods, costs for repositioning (energy costs, robot wear) are reasonable and charging times allow it, active repositioning can make a reasonable contribution to a system's overall performance. Since the introduced mechanisms greedily search for repositioning moves, more moves are conducted than necessary to obtain a desired inventory well-sortedness. For future research we suggest to predetermine moves before starting repositioning operations, e.g. by using a MIP formulation matching pods with storage locations and selecting the best moves. The source-code of this publication is available at \url{https://github.com/merschformann/RAWSim-O}.

\bibliographystyle{plain}
\bibliography{repositioning-preprint}

%

\end{document}